\documentclass[11pt,twoside]{article}

\usepackage{asp2006}
\usepackage{epsf}
\usepackage{psfig}
\usepackage{lscape}
\usepackage{graphicx}
\usepackage{sidecap}
\usepackage{wrapfig}

\begin{document}

\title{The Near-Star Environment: Spectropolarimetry of Herbig Ae/Be Stars}   
\author{D.M. Harrington \& J.R. Kuhn}   
\affil{University of Hawaii, Institute for Astronomy, 2680 Woodlawn Drive, Honolulu HI 96822}    

\begin{abstract} 

     The near-star environment around young stars is very dynamic with winds, disks, and outflows.  These processes are involved in star and planet formation, and influence the formation and habitability of planets around host stars.  Even for the closest young stars, this will not be imaged even after the completion of the next generation of telescopes decades from now and other proxies must be used.  The polarization of light across individual spectral lines is such a proxy that contains information about the geometry and density of circumstellar material on these small spatial scales.  We have recently built a high-resolution spectropolarimeter (R~13000 to 50000) for the HiVIS spectrograph on the 3.67m AEOS telescope.  We used this instrument to monitor several young intermediate-mass stars over many nights.  These observations show clear spectropolarimetric signatures typically centered on absorptive components of the spectral lines, with some signatures variable in time.  The survey also confirms the large spectroscopic variability in these stars on timescales of minutes to months, and shows the dyamic ÒbulletsÓ and ÒstreamersÓ in the stellar winds.  These observations were largely inconsistent with the traditional scattering models and inspired the development of a new explanation of their polarization, based on optical-pumping, that has the potential to provide direct measurements of the circumstellar gas properties.  

\end{abstract}

\section{Introduction}

	The formation of planets around stars is a dynamic process.  Over the course of a few million years, the circumstellar gas and dust will either accrete onto the star, turn in to planets, or dissipate in the form of winds and jets.  The star also evolves from an active young star showing emission lines (TTauri or Herbig Ae/Be stars) to a stable main-sequence star.  There is evidence of accretion, outflows, ionized disk structures, and strong stellar winds in young stars, many times happening simultaneously.  All of these processes influence the environment of the planets around these stars.  The star's physical properties influence the habitable zone and its planets' atmospheres through planet-star interactions.  By measuring properties of the circumstellar material in young stars, we can put direct constraints on these processes.  

	There are only a few techniques that can put meaningful constraints on the environment around a star.  Spectropolarimetry is one, but it is still considered a difficult specialty field.  In the stellar astronomical community, only a handful of spectropolarimeters exist.  There are only two high-resolution instruments, ESPaDOnS and HiVIS on large telescopes (over 3m).  However, this technique is a powerful probe of small spatial scales, being sensitive to the geometry and density of the material very close the central star.  Typically, material in the region a few to several stellar radii away from the star produces the polarization.  Even for the closest young stars (150pc), these spatial scales are smaller than 0.1 milliarcseconds across and will not be imaged directly, even by 100m telescopes.  Since the circumstellar material is involved in accretion, outflows, winds and disks, with many of these phenomena happening simultaneously, spectropolarimetry can put unique constraints on the types of densities and geometries of the material involved in these processes.  

\begin{figure}
\includegraphics[width=0.85\textwidth, height=1.1\textwidth, angle=90]{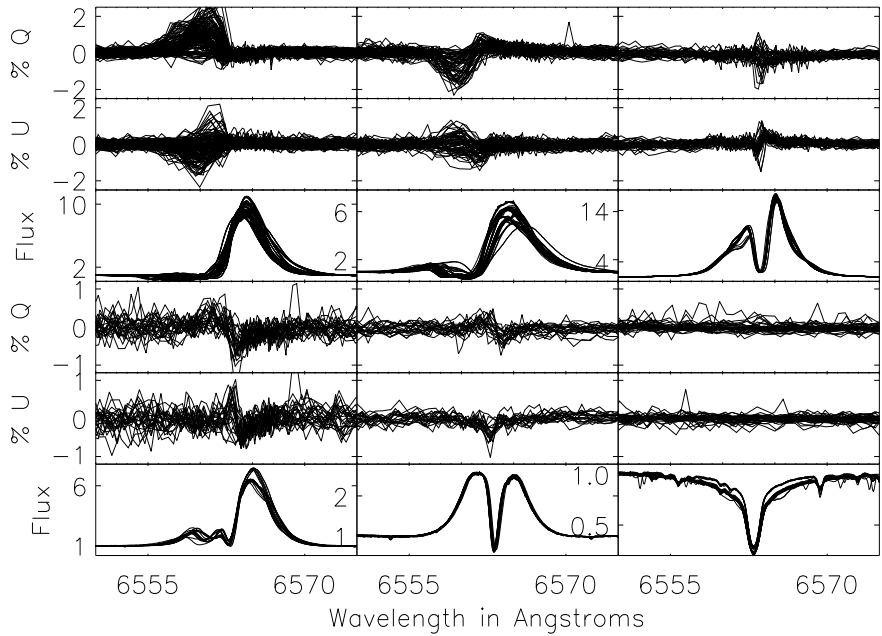}
\vspace{-30mm}
\caption[swap-all]{\label{fig:swap-all}
The spectroscopy and spectropolarimetry of 5 Herbig AeBe's and unpolarized stars in the lower right for comparison.  Each star has three boxes.  The left column is AB Aurigae above and MWC120 below.  The middle column is MWC480 and HD58647.  The right column, is MWC158, and unpolarized calibrators.  Stokes q, the vertical (+) and horizontal (-) polarization in percent in the top box.  Stokes u, the $\pm45^\circ$ polarization in percent is in the middle box. The average nightly spectra are in the bottom box.  See Harrington \& Kuhn 2007 for details. }
\end{figure}
	
	Spectropolarimetry in it's most simple form is just spectroscopy through a polarizer.  Exact instrumental techniques vary, but they all measure intensity {\it and} polarization as a function of wavelength.  With a high-resolution instrument ($\frac{\lambda}{\delta\lambda}\ge$15000) clear polarization changes across individual emission and absorption lines are seen.  Things like circumstellar disks, rotationally distorted winds, magnetic fields, or asymmetric radiation fields (optical pumping) can cause these signatures.  The polarization comes directly from the circumstellar material and can be used as a proxy for the physical properties of the circumstellar material, just like spectroscopy.  Typical spectropolarimetric signals are small, often a few tenths of a percent change in polarization.  Measuring these signals requires very high signal to noise observations and careful control of systematics errors.  We have recently built and calibrated a polarimetry package for the HiVIS spectrograph on the 3.67m AEOS telescope on Haleakala, HI to address these important issues (Harrington et al. 2006, Harrington \& Kuhn 2008).
	
	There are many models of spectropolarimetric effects.  Early analytical studies showed the possibility of spectropolarimetric effects from scattering very close to the central star (McLean 1979, Wood et al. 1993 \& 1994).  Recent modelling of scattering by circumstellar materials has shown a wealth of possible polarimetric line-effects from disks, winds, and envelopes (Harries 2000, Ignace 2004, Vink et al. 2005a).  Unpolarized line emission that forms over broad stellar envelopes can produce a depolarization in the line core relative to the stellar continuum.  Small clumps in a stellar wind that scatter and polarize significant amounts of light can enhance the polarization at that clump's specific velocity and orientation.  Magnetic fields can, through the Zeeman effect, shift the wavelength of polarized atomic emission, giving rise to a field-dependent polarization.  Optically pumped gas can produce polarization through absorption (Kuhn et al. 2007).

\section{Spectropolarimetry}

\begin{wrapfigure}{R}{0.5\textwidth}
\begin{center}
\vspace{-20mm}
\includegraphics[width=0.45\textwidth, height=0.65\linewidth]{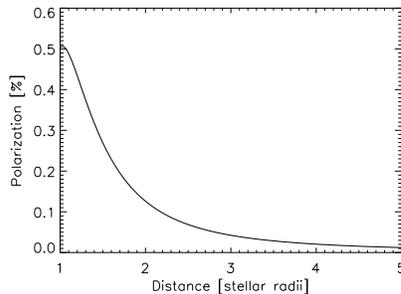}
\vspace{0mm}
\caption[pump]{ Polarization as a function of distance between a gas cloud and the star assuming a T=10,000K star with limb darkening.  The intervening cloud is obscuring the edge of the stellar disk.  See Kuhn et al. 2007 for details. \label{fig:pump}}
\end{center}
\end{wrapfigure}

	To date, only a few detections of spectropolarimetric signals in young stars have been reported, and the variability of these signatures has not yet been studied in detail (Vink et al. 2002, 2005b  Mottram et al. 2007, Harrington \& Kuhn 2007).  Our study, as well as most others, have focused on the H$_\alpha$ line, a strong line of Hydrogen.  We initially chose bright, well-known stars for close study but have since expanded our target list to include many young stars.  The H$_\alpha$ line in these stars is typically a very strong emission line with additional absorptive components.  The line can be up to 20 times continuum, typically with evidence for winds or disks: P-Cygni absorption profiles or central absorption components respectively.  Our observations of windy-stars showed strong variability of the absorption component, often over 10-minute timescales.  In disky stars, the H$_\alpha$ line showed strong central absorptions but were much less variable.  Modeling the spectra and variability alone can provide information about the near-star environment (cf. Beskrovnaya et al. 1995,  Bouret \& Catala 1998, Catala et al. 1999).
	  	
	Preliminary studies at medium resolution showed many different signatures with amplitudes up to 2\% change (Vink et al. 2002, 2005b).  Our observations of the H$_\alpha$ line for our first five stars, shown in figure \ref{fig:swap-all}, show similar amplitudes.  Instrumental effects cause a lot of the aparant variability in this figure, but there is intrinsic variability as well (see Harrington \& Kuhn 2008).  The signatures are roughly 0.3\% to 1.5\% centered on the absorption component, not on line center.  We have since observed polarization signatures that span the entire H$_\alpha$ line and have several non-detections, but the polarization-in-absorption is a common feature for intermediate-mass young stars.
	 
\section{A New Model - Optical Pumping}

	The simple scattering polarization and depolarization models all predicted signatures centered on the line core (McLean 1979, Wood et al. 1993, Vink et al. 2005a).  In most of our stars, the change in polarization occurred in and around the absorptive component, whether central or blue-shifted.  The polarization of the emission peak was nearly identical to the continuum polarization.  This problem led us to explore alternative explanations that would require the absorbing material to also be the polarizing material.  We developed a new model where the stellar radiation causes the absorbing material to polarize the transmitted light (Kuhn et al. 2007).  The circumstellar gas is optically pumped by the strong anisotropic radiation from the star causing the absorbing gas to absorb different incident polarizations by different amounts.  The main difference between this optical pumping model and the scattering model is that only the absorbing material is responsible for the changing polarization, whereas the scattering models integrate scattered light from the entire circumstellar region with each part contributing to the polarization change.  An example of an optical pumping calculation is shown in figure \ref{fig:pump}.  This new model is simple and deterministic.  With some further development, this model can be used to put a direct constraint on the circumstellar gas conditions.  This in turn will clarify our understanding of the circumstellar environment and interactions.

\acknowledgements
This program was partially supported by the NSF AST-0123390 grant, the University of Hawaii and the AirForce Research Labs (AFRL).  This program also made use of the Simbad data base operated by CDS, Strasbourg, France.

\end{document}